\documentclass[envcountset]{llncs}
\usepackage{amsmath}
\usepackage{amssymb}
\usepackage{enumerate}
\usepackage{graphicx} 
\usepackage{array}
\usepackage{arydshln}

\usepackage{tikz}
\usetikzlibrary{positioning}
\usetikzlibrary{arrows}

\newcommand{\problemFont}[1]{\protect\ensuremath{\mathsf{#1}}}
\newcolumntype{L}[1]{>{\raggedright\let\newline\\\arraybackslash\hspace{0pt}}m{#1}}
\newcolumntype{C}[1]{>{\centering\let\newline\\\arraybackslash\hspace{0pt}}m{#1}}
\newcolumntype{R}[1]{>{\raggedleft\let\newline\\\arraybackslash\hspace{0pt}}m{#1}}

\newcommand{\PTIME}{\problemFont{PTIME}}
\newcommand{\si}{\sigma}
\newcommand{\Si}{\Sigma}
\newcommand{\sub}{\subseteq}
\newcommand{\pr}{\mathrm{pr}}
\newcommand{\tuple}[1]{\vec{#1}}

\newcommand{\Dom}{\textrm{Dom}}

\newcommand{\N}{\mathbb{N}}

\newcommand{\Po}{\mathcal{P}}

\newcommand{\on}{\exists}
\newcommand{\ja}{\wedge}

\newcommand{\bow}[1]{\bowtie\hspace{-.6mm}(#1)}
\newcommand{\bo}{\bowtie}

\newcommand{\at}{\problemFont{Att}}

\newcommand\re[2]{{% we make the whole thing an ordinary symbol
  \left.\kern-\nulldelimiterspace % automatically resize the bar with \right
  #1 % the function
  \vphantom{|} % pretend it's a little taller at normal size
  \right|_{#2} % this is the delimiter
  }}
\newcommand{\Val}{\problemFont{Val}}
\newcommand{\chase}[1]{\overline{(#1)}}
\newcommand{\ochase}[1]{\textrm{chase}\overline{(#1)}}
\newcommand{\id}{\mathrm{id}}
\spnewtheorem{ex}{Example}{\bfseries}{\rmfamily}

\begin{document}

\title{Reasoning about embedded dependencies\\ using inclusion dependencies}
\author{Miika Hannula}
\institute{University of Helsinki, Department of Mathematics and Statistics,
P.O. Box 68, 00014 Helsinki, Finland \email{miika.hannula@helsinki.fi}}
\maketitle
\

\begin{abstract}
The implication problem for the class of embedded dependencies is undecidable. However, this does not imply lackness of a proof procedure as exemplified by the chase algorithm. In this paper we present a complete axiomatization of embedded dependencies that is based on the chase and uses  inclusion dependencies and implicit existential quantification in the intermediate steps of deductions. %In the system, intermediate steps of deductions are inclusion dependencies that introduce new attributes, thought of as implicitly existentially quantified. 
%These inclusion dependencies may introduce so-called new attributes which can be thought of as implicitly existentially quantified.
 \end{abstract}


\section{Introduction}
Embedded dependencies generalize the concept database dependencies within the framework of first-order logic. %Usually dependencies express that the presence of certain tuples in a relation implies presence of other tuples or equality of some tuple components. Since expressions of this sort can be formalized in first-order logic, it is therefore in principle possible to axiomatize such expressions. 
Their implication is undecidable but however recursively enumerable, thus enabling complete axiomatizations. A standard example of such a proof procedure is the chase that was invented in the late 1970s \cite{aho79,maier79}, and then soon  extended to  equality and tuple generating dependencies \cite{beeri84}. In this paper we present an axiomatization for the class of embedded dependencies that simulates the chase at the logical level 
using inclusion dependencies. In particular, completeness of the rules is obtained by constructing deductions in which all the intermediate steps are inclusion dependencies, except for the first and the last step. These inclusion dependencies consist of attributes of which some are new, i.e., such that they are not allowed to appear at any earlier stage of the deduction. 

As a background example, consider the combined class of functional and inclusion dependencies. It is well known that the corresponding implication problem is undecidable, lacking hence finite axiomatization \cite{chandra85,mitchell83}.  In these situations, one strategy has been to search for axiomatizations within a more general class of   dependencies, and partly for this reason many different dependency notions were introduced in the 1980s. For instance, a textbook on dependency theory from 1991 considers more than  80 different dependency classes \cite{thalheim91}. In \cite{mitchell83b} Mitchell proposed  another strategy by presenting an axiomatization of functional and inclusion dependencies using a notion of new attributes which are to be thought of as implicitly existentially quantified. %In this system the associated proof procedures did not necessarily terminate because each deduction is allowed contain an unbounded number of attributes.
%. It is well known that implication in their combined class is undecidable \cite{chandra85,mitchell83}, and lacks hence finite axiomatization in the usual sense. A common strategy in these situations has been to search for axiomatization within a more general class of dependencies. Another presented by Mitchell enables finite axiomatizations within the class itself, since the proof procedures do not necessarily terminate because of  the possibility of introducing new attributes.
In this paper we take an analogous approach, and present an axiomatization for embedded dependencies where new attributes correspond to new  values obtained from an associated chasing sequence. %At the logical level, t
These attributes are implicitly existentially quantified in the sense of \emph{team semantics}, that is, a semantic framework that has \emph{teams}, i.e.,  sets of assignments, as its underlying concept \cite{hodges97}. Team semantics is compositionally applicable to logics that extend first-order logic with various database dependencies \cite{galliani12,vaananen07}.  %In expressive power such logics lie between first-order logic and existential second-order  logic (\ESO). For instance, 
In this setting, %\emph{dependence logic} obtained by adding functional dependencies to first-order logic captures $\ESO$ in expressive power, and %, and hence all $\NP$ recognizable classes of finite models 
\emph{inclusion logic}, i.e., first-order logic with additional inclusion dependencies, captures positive greatest fixed-point logic and hence all $\PTIME$ recognizable classes of finite, ordered models \cite{gallhella13,immerman86,vardi82}. Therefore, inclusion dependencies with new attributes can be  thought of as existentially quantified inclusion logic formulae which in turn translate into greatest fixed-point logic. Moreover, all existentially quantified dependencies that appear in deductions translate into existential second-order logic. This may in part enable succinct intermediate  steps in contrast to axiomatic systems that simulate the chase by composing first-order definable dependencies. 

The methods described in this paper generalize the axiomatization of conditional independence and inclusion dependencies presented in \cite{hannula14}. It is also worth noting that extending relations with new attributes reminds of algebraic dependencies, that are, typed embedded dependencies defined in algebraic terms. %, having their semantics defined using horizontally extended relations. 
The complete axiomatization of algebraic dependencies presented in \cite{yannakakis82} involves also an extension schema that introduces new copies of attributes.
%In contrast to the usual Tarskian semantics of first-order logic, 
%Team semantics connects  to relational databases since it has sets of assignments as its underlying concept. Regarding our example, first-order logic with additional functional dependencies (or inclusion dependencies) is called dependence (inclusion) logic \cite{vaananen07,galliani12}. Interestingly, inclusion logic over finite ordered models captures $\PTIME$ whereas dependence logic over finite models captures $\NP$ \cite{vaananen07,gallhella13}. This is in contrast with the fact that the implication problem for functional dependencies alone is much easier to compute than that for inclusion dependencies (the first is in linear time and the second $\PSPACE$-complete \cite{beeri79,casanova82}).

%Existential quantification in (lax) team semantics extends each row of a team with a set of values, thus possibly resulting in a blow up in the cardinality of the team.

%A \emph{partial tuple} over $R$ is a partial mapping $R\to\Val$, and a \emph{partial relation} over $R$ is a set of partial tuples over $R$. We define values, valuations, extension regarding partial tuples and  relations analogously to the previous paragraph. Let $r$ and $r'$ be a partial and a total relation, respectively. Then we write  $r \leq r'$ if for all $t\in r$ there exists $t'\in r'$ such that $t(A)=t'(A)$ for all $A \in \Dom(t)$. If $f$ is a valuation such that $f(r) \leq r'$, then we say that $f$ embeds $r$ to $r'$.

\section{Preliminaries}
%We start by fixing two countably infinite sets $\Var$ and $\at$, the first denoting variables and the second  attributes. 
For two sets $A$ and $B$, we write $AB$ to denote their union, and for two sequences $\tuple a \tuple b$, we write $\tuple a \tuple b$ to denote their concatenation. %or the sequence $(A,B)$, depending on the context. 
For a sequence $\tuple a =(a_1, \ldots ,a_n)$ and a mapping $f$, we write $f(\tuple a)$ for $(f(a_1),\ldots ,f(a_n))$. We denote by $\id$ the identity function and by $\pr_i$ the function that maps a sequence to its $i$th projection. 
 %For two sequences $\tuple a$ and $\tuple b$, we write $\tuple a\tuple b$ for their concatenation. For two sets $A$ and $B$, we write $AB$ for their union. 
For a function $f$ and $A\sub \Dom(f)$, we write $\re{f}{A}$ for the restriction of $f$ to $A$, and for a set of mappings $F$, we write $\re{F}{A}$ for $\{\re{f}{A}: f\in F\}$.

We start by fixing two countably infinite sets $\Val$ and $\at$, the first denoting possible values of relations and the second  attributes. For notational convenience, we will assume that $\Val= \at$. For $R\sub \at$, a \emph{tuple} over $R$ is a  mapping $R \to \Val$, and a \emph{relation} over $R$ is a set of tuples over $R$. We may sometimes write $r[R]$ to denote that $r$ is a relation over $R$. Values of a relation $r$ over $ R$ are denoted by $\Val(r)$, i.e., $\Val(r):=\{t(A):t\in r,A\in R\}$. Let $f$ be a \emph{valuation}, i.e., a mapping $\Val \to \Val$. % extend to relations and dependencies as follows. %For a tuple $t$, $f(t):= f\circ t$, and 
Then for a tuple $t$, we write $f(t):=f\circ t$, and for a relation $r$, $f(r) :=\{f( t) :t\in r\}$. A valuation $f$ \emph{embeds}  a relation $r$ (a tuple $t$) to $r'$ if $f(r) \sub r'$ ($f(t)\in r$). Since we are usually interested only valuations of a relation, we say that $f: \Val(r)\to \Val$ is a \emph{valuation on $r$}. For a valuation $f$ on $r$, we say that $g$ is an \emph{extension} of $f$ to another relation $r'$ if $g$ is a valuation on $r'$ such that it agrees with $f$ on values of $\Val(r)\cap\Val(r')$.

%LISÄÄ valuation/extension on tuples !!! ks. chase soundness todistus
\emph{Embedded dependencies} (ed's) can be written using first-order logic in the following way.
\begin{definition}[Embedded dependency]
Embedded dependency is a first-order sentence of the form
$$\forall x_1, \ldots ,x_n \big ( \phi(x_1, \ldots, x_n) \rightarrow \on z_1 \ldots \on z_k\psi(y_1, \ldots ,y_m)\big )$$
where $\{z_1, \ldots ,z_k\} = \{y_1, \ldots ,y_m\}\setminus \{x_1, \ldots ,x_n\}$ and
\begin{itemize}
\item $\phi$ is a (possibly empty) conjunction of relational atoms using all of the variables $x_1, \ldots ,x_n$;
\item $\psi$ is a conjunction of relational and equality atoms using all of the variables $z_1, \ldots ,z_k$; 
\item there are no equality atoms in $\psi$ involving existentially quantified variables.
\end{itemize}

\end{definition}
If at most one relation symbol occurs in an ed, then we say that the ed is \emph{unirelational}, and otherwise it is \emph{multirelational}. An ed is called \emph{typed} if there is an assignment of variables to column positions such that variables in relation atoms occur only in their assigned position, and each equality atom involves a pair of variables assigned to the same position. Otherwise we say that an ed is \emph{untyped}. If $\psi$ contains only one atom, then we say that the ed is \emph{single-head}, and otherwise it is \emph{multi-head}.
A single-head ed where $\psi$ is an equality is called an \emph{equality generating dependency} (egd). If $\psi$ is a conjunction of relational atoms, then the ed is called a \emph{tuple generating dependency} (tgd). For notational simplicity, we restrict attention to unirelational ed's. It is easy to se that any ed is equivalent to a set of tgd's and egd's, and hence we restrict attention to ed's that belong to either of these subclasses.

The following alternative tableau presentation for egd's and tgd's are used in this paper. %Regarding the first  definition note that those values that occur exactly once in $T\cup T'$ do not play any role. Hence we may also define egd's and tgd's using partial relations.
\begin{definition}
Let $T$ and $T'$ be finite relations over $ R$, and $x,y\in \Val(T)$. Then $(T,x=y)$ and $(T,T')$ are an egd and a tgd over $R$, respectively, with the below satisfaction relation for a relation  $r$ over $S\supseteq R$:
\begin{itemize}
\item $r \models (T, x =  y) \Leftrightarrow$  for all valuations $f$ such that $f(T) \sub \re{r}{R}$, it holds that  $f(x)=f(y)$.
\item $r \models (T,T') \Leftrightarrow$   for all valuations $f$ on $T$ such that $f(T)\sub \re{r}{R}$, there is an extension $g$ of $f$ to $T'$ such that $g(T')\sub \re{r}{R}$. 
\end{itemize}
\end{definition}
Sometimes we write $\si[R]$ to denote that $\si$ is a dependency over $R$. 
If $T$ or $T'$ is a singleton, then we may omit the set braces in the notation, e.g., write $(T,t)$ instead of $(T,\{t\})$. 

We also extend valuations to dependencies. For an egd $\si= (T,x=y)$ we write $\Val(\si)=\Val(T)$, and for a tgd $\tau= (T,T')$  we write $\Val(\si)=\Val(T)\cup\Val(T')$. Moreover,  if $f$ is a valuation, then $f(\si)=(f(T),f(x)=f(y))$ and $f(\tau)=(f(T),f(T'))$.
%If $T$ and $T'$ are relations over $R$, then we may use $(T,T')$ (or $(T,x=y)$) to denote $(T,T')[R]$ (or $(T,x=y)[R]$).
%When presenting relations as  lists of its tuples, we  omit the set braces in the notation, e.g., we write $(t,t',u)$ instead of $(\{t,t'\},\{u\})$ and say that $f$ embeds $t,t'$ (instead of $\{t,t'\}$) into $r$. 

\begin{ex}
Consider the relation $r$ and the tgd's $\sigma_1:= (\{t,t'\},\{u\})$ and $\sigma_2:= (\{t,t'\},\{v,v'\})$ obtained from Fig. \ref{kuv}.\footnote{In a tableau presentation of a dependency $\si$, the distinct values of $\si$ are sometimes denoted by blank cells.}
%&$\si=(T,T')$ (or $\si=(T, x =  y)$), we may omit the distinct values of $\Val(T) \cup \Val(T')$ (or of $\Val(T)$), i.e., the values that appear at most once in $T\cup T'$ (or in $T$ and $x,y$).}
%  and present the new attributes of the rule on the right-hand side of the vertical dashed line.}  
We notice that there are two valuations on $\{t,t'\}$  that embed $\{t,t'\}$ to $r$, namely $f:=\{(x,0),(y,1),(z,2)\}$ and $g:=\{(x,3),(y,0),(z,1)\}$. Then $r\models \si_1$ since $f$ and $g$ embed $u$ into $r$, witnessed by tuples $s_2$ and $s_3$, respectively. We also notice that $r\not\models \si_2$ since, although $f\cup\{(a,3)\}$ embeds $\{v,v'\}$ into $r$, no extension of $g$ does the same.
\end{ex}

\begin{figure}[h]
\begin{center}
$ \scalebox{1.2}[1.1]{$r=$ \begin{tabular}{ c | C{3.5mm} C{3.5mm} C{3.5mm} | }
&$A$& $B$ & $C$\\\hline
$s_0$&$0$ & $1 $ & $2$ \\\hline
$s_1$&$3$ &$0 $ & $1$\\\hline
$s_2$&$2$ &$ 3$ & $0$\\\hline
$s_3$&$1$ &$ 4$ & $3$\\\hline

\end{tabular}

\qquad

$\sigma_1= $ \begin{tabular}{ c | C{3.5mm} C{3.5mm} C{3.5mm}  | }
&$A$& $B$ & $C$\\\hline
$t$&$x$ & $y $ & $z$ \\
$t'$& &$x $ & $y$\\\hline
$u$&$z$ & & $x$\\\hline

\end{tabular}
\qquad

$\sigma_2= $ \begin{tabular}{ c | C{3.5mm} C{3.5mm} C{3.5mm}  | }
&$A$& $B$ & $C$\\\hline
$t$&$x$ & $y $ & $z$ \\
$t'$& &$x $ & $y$\\\hline
$v$&$z$ &$a$ & $x$\\
$v'$& & & $a$\\\hline

\end{tabular}

}$\caption{\label{kuv}}

\end{center}
\end{figure}

%In the following axiomatization, some rules introduce new attributes which extend horizontally the size of a relation. Hence we relativize our presentation of tgd's and egd's  as follows. Let $T$ and $T'$ be partial mappings $\at \to \Var$, and let  $r$ be a relation over $S \supseteq R$. Then we write $r \models (T, x =  y)[R]$ (or $r \models (T, T')[R]$) iff $\re{r}{R} \models (\re{T}{R}, x =  y)$ (or $\re{r}{R} \models (\re{T}{R},\re{T'}{R})$. 
 
Next we define inclusion dependencies which are examples of possibly untyped tgd's. 
\begin{definition}[Inclusion dependency]
Let $A_1, \ldots ,A_n$ and $B_1, \ldots ,B_n$ be (not necessarily distinct) tuples of attributes. Then $A_1 \ldots A_n \sub B_1 \ldots B_n$ is an \emph{inclusion dependency} (ind) over $R= \{A_i,B_i: i=1, \ldots ,n\}$ with the following semantic rule for a relation $r$ over $S\supseteq R$:
$$r\models A_1 \ldots A_n \sub B_1 \ldots B_n \Leftrightarrow \forall s\in r\exists s'\in r \forall i=1, \ldots ,n: s(A_i)=s'(B_i).$$
\end{definition}
The axiomatization presented in the next section involves inclusion dependencies that introduce new attributes. These attributes are here interpreted as existentially quantified in \emph{lax team semantics} sense \cite{galliani12}:
\begin{equation}\label{lax}
r\models \on A \phi \Leftrightarrow r[f/ A] \models \phi \textrm{ for some } f:r\to \Po(\Val)\setminus\{\emptyset\},
\end{equation}
where $r[f/ A]:=\{t(x/A) : x\in f(A)\}$ and $t(x/A)$ is the mapping that agrees with $t$ everywhere except that it maps $A$ to $x$. Interestingly, inclusion logic formulae with this concept of existential quantification can be characterized with positive greatest fixed-point logic formulae (see Theorem 15 in \cite{gallhella13}).

%, i.e., such that some value appears in two distinct columns in the tableau presentation. An egd is called untyped if the previous condition holds or the associated equality involves values that appear in two distinct columns. %The implication problem of ind's alone enjoys finite axiomatization and is $\PSPACE$-complete [\cite{chandra85}]. 

%If $f$ and $g$ are functions from $X$ to attributes, then we write $f(X) \sub g(X)$ to denote any $f(X_1) \ldots f(X_n) \sub g(X_1)\ldots g(X_n)$ where $X_1, \ldots ,X_n$ is an enumeration of $X$. We also write $f(X) \sub X$ and $X \sub f(X)$ for $f(X) \sub \textrm{id}(X)$ and $ \textrm{id}(X)\sub f(X)$, respectively. 

\section{Axiomatization}\label{axiomatization}
In this section we present an axiomatization for the class of all embedded dependencies. The axiomatization contains an identity rule and three rules for the chase. We also involve conjunction in the language and therefore incorporate its usual introduction and elimination rules in the definition.  
%It is straightforward to show that every ed can be expressed as a conjunction of tgds and \emph{single-head} egds. Hence and for notational simplicity, we present this axiomatization for the class of tgds and single head egds. %Also 
Regarding the equalities that appear in the rules, note that both $AB\sub AA$ and $AB \sub BB$  indicate that the values of $A$ and $B$ coincide in each row. Therefore, we use $A=B$ to denote ind's of either form. For a tgd (an egd) $\si$, we say that $x\in \Val(\si)$ is \emph{distinct} if it appears at most once as a value in $\si$. Namely, 
\begin{itemize}
\item for a tgd $\si=(T,T')[R]$, $x$ is distinct if for all $t,t'\in T\cup T'$ and $A,B\in R$, if $t(A)=x=t'(B)$, then $t=t'$ and $B=B'$;
\item for an egd $\si=(T,y=z)[R]$, $x$ is distinct if $x\not\in\{y,z\}$ and for all $t,t'\in T$ and $A,B\in R$, if $t(A)=x=t'(B)$, then $t=t'$ and $B=B'$.

\end{itemize}
%We call a set of distinct values also distinct. 
Lastly, note that in the following  rules we  assume that values can appear as attributes and vice versa.
%Hence, in the following axiomatization we think of any rule involving equalities or ind's of the form $f(R)\sub g(R)$ as a rule schema consisting of suitable instances. Note also that instances of $A=B$ (or $f(R)\sub g(R)$) are equivalent and derivable from one another using rules [EE,IP]. 

\begin{definition}\label{axioms}
In addition to the below rules we adopt the usual introduction and elimination rules for conjunction. In the last three rules, we assume that $\tuple A$ is a sequence listing the attributes of $R$.
\\
%\textbf{Rule for ind's}:
\begin{itemize}
%\item[IR] Inclusion Reflexivity:   $$\vec{A} \subseteq \vec{A}.$$

%\item[IP] Inclusion Projection and Permutation: $$\textrm{if }A_1 \ldots A_n \subseteq B_1 \ldots B_n  \textrm{, then } A_{i_1} \ldots A_{i_k} \subseteq B_{i_1} \ldots B_{i_k},$$ for each sequence $i_1 , \ldots ,i_k$ of integers from $\{1, \ldots ,n\}$.
%\item[IT] Inclusion Transitivity: $$\textrm{if }\vec{A} \subseteq \vec{B}\ja \vec{B} \subseteq \vec{C} \textrm{, then }\vec{A} \subseteq \vec{C}.$$

%\item[II] Inclusion Introduction: $$\textrm{if }\vec{A} \subseteq \vec{B} \textrm{, then } \vec{A}E \subseteq \vec{B} C,$$ where $E$ is a \emph{new} attribute from $\ex$.

\item[EE] Equality Exchange: $$\textrm{if }A=B  \ja \sigma \textrm{, then } \tau.$$
where $\si$ is an ind and $\tau$ is obtained from $\si$ by replacing any number of occurrences of $A$ by $B$ and any number of occurrences of $B$ by $A$.
%\item[SI] Split Introduction: $$\tuple E = \tuple F$$
%\end{itemize}
%\textbf{Rules for chase}:
%\begin{itemize}
%\item[CS] Chase Start: $$(T,u)[RS]\ja \bigwedge_{t\in T}  u^{-1}\circ t(\tuple A) \sub  \tuple A$$

%where $\tuple A$ is a sequence listing the attributes of $R$, $S$ is a set of \emph{new} attributes, $\re{u}{S}$ is bijection $S\to \Val(\re{T}{R})$, and $\Val(\re{u}{R})$ is distinct. 

%\item[CR] Chase Rule: 
%$$\textrm{\hspace{0cm}tgd:\quad}\textrm{if }( T, T')[R] \ja \bigwedge_{t\in T}f\circ t(\tuple A) \sub \tuple  A \textrm{, then } \bigwedge_{t'\in T'}f\circ t'(\tuple A) \sub \tuple A,$$
%$$\hspace{0cm}\textrm{egd:\quad}\textrm{if }(T,x=y)[R] \ja \bigwedge_{t\in T}  f\circ t(\tuple A) \sub \tuple A \textrm{, then } f(x)=f(y),$$

%where $\tuple A$ is a sequence listing the attributes of $R$, $f$ is a function $\Val\to \at$, and $\Val(f(T')) \setminus \Val(f(T))$ is a set of  \emph{new} attributes.

%\item[CT] Chase Termination: 
%$$\textrm{\hspace{0cm} tgd:\quad}\textrm{if }(T,u)[R S]\ja\bigwedge_{t'\in T'} u'\circ t'(\tuple A)\sub \tuple A
%,\textrm{ then } (\re{T}{R} ,T')[ R],$$
%$$\textrm{\hspace{0cm} egd:\quad}\textrm{if }(T,u)[ R  S]\ja u^{-1}(x)=u^{-1}(y)
%,\textrm{ then } (\re{T}{R} ,x=y)[R],$$
%where  $\tuple A$ is a sequence listing the attributes of $R$, $\re{u}{S}$ is bijection $ S\to \Val(\re{T}{R})$,  and $\Val(\re{T}{S})$ is distinct. Moreover, tgd: $u'$ is a mapping  $\Val(T')\to \at$ that agrees with  $u^{-1}$ on $\Val(T')\cap\Val(\re{T}{R})$, and egd: $x,y\in \Val(\re{T}{R})$.
%\end{itemize}
%\begin{itemize}
\item[CS] Chase Start: $$(T^*,\id)[RS]\ja \bigwedge_{t\in T}   t(\tuple A) \sub  \tuple A$$

where $T=\re{T^*}{R}$, $S= \Val(T)$  consists of \emph{new attributes}, and $R$ consists of \emph{distinct values}. 

\item[CR] Chase Rule: 
$$\textrm{\hspace{0cm}tgd:\quad}\textrm{if }( T, T')[R] \ja \bigwedge_{t\in T}f\circ  t(\tuple A) \sub \tuple  A \textrm{, then } \bigwedge_{t'\in T'}f\circ  t'(\tuple A) \sub \tuple A,$$
$$\hspace{0cm}\textrm{egd:\quad}\textrm{if }(T,x=y)[R] \ja \bigwedge_{t\in T} f\circ  t(\tuple A) \sub \tuple A \textrm{, then } f(x)=f(y),$$

where tgd: $f$ is a valuation that it is 1-1 on $\Val(T')\setminus\Val(T)$, and $f(x)$ is a \emph{new attribute} for $x\in \Val(T')\setminus\Val(T)$.

\item[CT] Chase Termination: 
$$\textrm{\hspace{0cm} tgd:\quad}\textrm{if }(T^*,\id)[R S]\ja\bigwedge_{t'\in T'} u\circ t'(\tuple A)\sub \tuple A
,\textrm{ then } (T ,T')[ R],$$
$$\textrm{\hspace{0cm} egd:\quad}\textrm{if }(T^*,\id)[ R  S]\ja x=y
,\textrm{ then } (T ,x=y)[R],$$
%where  $S= \Val(\re{T}{R})$,  and $\Val(\re{T}{S})$ consists of \emph{distinct values}. Moreover, tgd: $u$ is a mapping  $\Val(T')\to \at$ that is the identity on $\Val(T')\cap\Val(\re{T}{R})$, and egd: $x,y\in \Val(\re{T}{R})$.\\

where  $T=\re{T^*}{R}$, $S= \Val(T)$, and $\Val(\re{T^*}{S})$ consists of \emph{distinct values}. Moreover, tgd: $u$ is a mapping  $\Val(T')\to \at$ that is the identity on $\Val(T)\cap\Val(T')$, and egd: $x,y\in \Val(T)$.
\end{itemize}
\end{definition}

For a dependency $\si$ over $R$, we let $\at(\si):=R$, and for a set of dependencies $\Si$, we let $\at(\Si):=\bigcup_{\si\in\Si}\at(\si)$ .

%In an application of Inclusion Introduction, the variable $x$ is called the new variable of the deduction step. Similarly, in an application of Start Axiom, the variables of $\vec{x}$ are called the new variables of the deduction step.
\begin{definition}\label{dedu}
A deduction from $\Si$ is a sequence $(\si_1, \ldots ,\si_n)$ such that:
\begin{enumerate}
\item Each $\si_i$ is either an element of $\Si$, an instance of [CS], or follows from one or more formulae of $\{\sigma_1, \ldots ,\sigma_{i-1}\}$ by one of the rules presented above.
\item For each $A\in \at(\si_i)$, if $A$ is new in $\si_i$, %where $\si_i$ is an instance of [CS] (or follows from $\{\sigma_1, \ldots ,\sigma_{i-1}\}$ by a tgd [CR]), 
then $A\not\in \at(\Si\cup\{\sigma_1,\ldots ,\sigma_{i-1}\})$, and otherwise $A\in \at(\Si\cup\{\sigma_1,\ldots ,\sigma_{i-1}\})$.
%\item $\si_i = \Var(\Si_i)\cap \ex \bo \Var(\Si_i) \cap \at$ and $\tau$ is a conjunction of equalities $E = F$ where $EF \sub \Var(\S_i) \cap \ex$.
%\item No $\si_j$ with $j > i$ is obtained from a formula containing dependencies of the form $\bo (\tuple A_i)_i$ where $\bigcup_i A_i \cap \ex \neq \emptyset$, or $\tuple A \sub \tuple B$ where $B \cap \ex \neq \emptyset$.
\end{enumerate}
We say that $\si$ is provable from $\Si$, written $\Si \vdash \si$, if there is a deduction $(\si_1, \ldots ,\si_n)$ from $\Si$ with $\si=\si_n$ and such that no attributes in $\si$ are new in $\si_1, \ldots ,\si_n$. %If $\phi$ is provable from the empty set, then we will write $\vdash \phi$.
\end{definition}

 We will also use the following rules that are derivable from [EE]:\\
%\textbf{Rules for equalities}:
\begin{itemize}
%\item[ER] Equality Reflexivity: $$A =A.$$
\item[ES] Equality Symmetry: $$\textrm{if }A=B\textrm{, then } B=A.$$
\item[ET] Equality Transitivity: $$\textrm{if }A = B \ja B = C \textrm{, then }A = C.$$
%\item[ES] Equality Symmetry: $$\textrm{if }A = B  \textrm{, then } B=A.$$

%\item[EI] Equality Introduction: $$\tuple E = \tuple F,$$
%where $\tuple E$ and $\tuple F$ are sequences of variables from $\mathcal{E}$.

%\item[EE] Equality Exchange: $$\textrm{if }AB \sub CC  \ja \sigma \textrm{, then } \tau.$$
%where $\tau$ is obtained from $\sigma$ by replacing all occurrences of $A$ by $B$.
\end{itemize}
One may find the chase rules  slightly convoluted at first sight. However, the ideas behind the rules are relatively simple as illustrated in the following examples.

%%%%%%%%%%%EXAMPLE 1
\begin{ex}[Chase Start]
%Consider $\si_0$ in Figure \ref{1}. 
Let $\si_0:= (\{t_0,t_1\},\{u_0\})[RS]$ be as in Figure \ref{3}, for $R:=\{A,B,C\}$ and $S:=\{x,y,z\}$.
\begin{figure}[h]
\center\scalebox{1.2}[1.1]{
$\si_0=\begin{tabular}{c | C{3.5mm}  C{3.5mm} C{3.5mm} : C{3.5mm}  C{3.5mm} C{3.5mm} |}
&$A$& $B$ & $C$ &$x$&$y$&$z$\\\cline{2-7}
$t_0$&$x$ & $y $ & $z$ &&&\\
$t_1$& &$x $ & $y$&&&\\\cline{2-7}
$u_0$&&&&$x$&$y$&$z$\\\cline{2-7}
\end{tabular}$%}%\caption{\label{1}}
\quad 
%\scalebox{1.2}[1.1]{
$\si_1=\begin{tabular}{c | C{3.5mm}  C{3.5mm} C{3.5mm} |}
&$A$& $B$ & $C$  \\\cline{2-4}
$t_0$&$x$ & $y $ & $z$  \\
$t_1$&&$x $ & $y$ \\\cline{2-4}
$u_1$&$z$ &  & $x$\\\cline{2-4}

\end{tabular}$
\quad 

$\si_2=\begin{tabular}{c | C{3.5mm}  C{3.5mm} C{3.5mm} |}
&$A$& $B$ & $C$  \\\cline{2-4}
$t_0$&$x$ & $y $ & $z$\\
$t_1$&   &  $x $ & $y$ \\\cline{2-4}
$u_2$&$z$&  & $v$ \\
$u_3$&$v$&  & $z$\\\cline{2-4}

\end{tabular}$
}\caption{\label{3}}
\end{figure}
Then
\begin{equation*}\label{eq-1}
\tau:=\si_0\ja xyz\sub ABC \ja xy\sub BC
\end{equation*}
is an instance of [CS].  Here  $x,y,z$ are interpreted either as  values or as new attributes. By the latter we intuitively mean that any relation $r[ABC]$ can be extended to some  $r'[ABCxyz]$ such that $r' \models \tau$. For instance, one can define $r':=q(r)$ where $q$ is the following SPJR  query
$$ ABC \bo (\pi_{xyz}(\si_{xy=BC}(\rho_{xyz/ABC}(ABC)\bo ABC)))$$
where $\si$ refers to (S)election, $\pi$ to (P)rojection, $\bo$ to (J)oin, and $\rho$ to (R)ename operator. Then $q(r)$ is a relation over $RS$ such that its restriction to $ xyz$ lists all $abc$ for which there exist $s,s'\in r$ such that $s(ABC)=abc$ and $s'(BC)=ab$. Let $\si_1=(\{t_0,t_1\},\{u_1\})[R]$ be as in Figure \ref{3}. 
Now,
\begin{equation*}\label{eq3}
r\models \si_1 \Leftrightarrow q(r)\models zx \sub AC.
\end{equation*}
%Moreover, $\Si \models \si_1$ if for all $r\models \Si$, $ q(r) \models  A_zA_x \sub AC$. 
Hence proving $\Si \models \si_1$ reduces to showing that $\Si\cup\{\tau\} \models zx\sub AC$.
\end{ex}

%EXAMPLE 2
\begin{ex}[Chase Rule]
 Assume 
\begin{equation}\label{eq2}
\si_2 \ja   xyz \sub ABC\ja  xy \sub BC
\end{equation}
where $\si_2= (\{t_0,t_1\},\{u_2,u_3\})[R]$ is as in Fig. \ref{3}, for $R:=\{A,B,C\}$. Then, interpreting $f$ as $\id$, one can derive with one application of [CR]   %Note that $R_{u_1}=R_{u_2}= \{AC\}$.
\begin{equation}\label{eq4}
zv \sub AC \ja vz \sub AC
\end{equation}
 from \eqref{eq2}. Note that in \eqref{eq4} $v$ is interpreted as a \emph{new} attribute, and the idea is that any relation $r[R]$  satisfying \eqref{eq2} and with  $v\not\in R$ can be extended to a relation $r'[R\cup\{v\}]$ satisfying \eqref{eq4}  by introducing  suitable values for $v$.
%\begin{figure}[h]
%\center\scalebox{1.2}[1.1]
%{
%$\si_2=\begin{tabular}{c | C{4mm}  C{4mm} C{4mm} |}
%&$A$& $B$ & $C$  \\\cline{2-4}
%$t_0$&$x$ & $y $ & $z$\\
%$t_1$&   &  $x $ & $y$ \\\cline{2-4}
%$u_2$&$z$&  & $v$ \\
%$u_3$&$v$&  & $z$\\\cline{2-4}
%\end{tabular}$}
%\caption{\label{2}}
%\end{figure}
\end{ex}

%EXAMPLE 3
\begin{ex}[Chase Termination]
Assume
\begin{equation}\label{eq3}
\si_0 \ja zx\sub AC
\end{equation}
where $\si_0= (\{t_0,t_1\},\{u_0\})[RS]$ is as in Fig. \ref{3}, for $R:=\{A,B,C\}$ and $S:=\{x,y,z\}$.
Then, letting $u=\id$, one can derive $\si_1$ as in Fig. \ref{3} from  \eqref{eq3} with one application of [CT].

%\begin{figure}
%\center
%$\si_2:=\begin{tabular}{c | c c c | c c c c c |}
%&$A$& $B$ & $C$ &$D_1$&$D_2$&$D_3$&$D_4$&$D_5$\\\cline{2-9}
%$t$&$x$ & $y $ & $z$ &$a_1$&$a_2$&$a_3$&$a_4$&$a_{13}$\\
%$t'$&$a_5$ &$x $ & $y$&$a_6$&$a_7$&$a_8$&$a_{9}$&$a_{14}$\\\cline{2-9}
%$s$&$a_{10}$ & $a_{11}$ & $a_{12}$&$x$&$y$&$z$&$a_{5}$& $a_{15}$\\\cline{2-9}
%\end{tabular}$\caption{\label{3}}
%\end{figure}

\end{ex}

\section{Soundness Theorem}
In this section we show that the axiomatization presented in the previous section is sound.  First note that the next lemma  follows from the definitions of egd's, tgd's and ind's.
\begin{lemma}\label{clear}
Let $\si $ be a dependency over $R$, and let $r$ and $r'$ be relations over supersets of $R$ and with $\re{r}{R}=\re{r'}{R}$. Then $r\models \si \Leftrightarrow r'\models \si$.
\end{lemma}
Then we prove the following lemma  which implies soundness of the axioms. For attribute sets $R,R'$ with $R\sub R'$ and a relation $r$ over $R$, we say that a relation $r'$ over $R'$ is an extension of $r$ to $R'$ if $\re{r'}{R}= r$. Recall from equation \ref{lax} that exactly such extensions are used in the existential quantification of lax team semantics.
\begin{lemma}\label{soundlem}
Let $r$ be a relation over $\at(\Si)$ such that $r\models \Si$, and  let $(\si_1, \ldots ,\si_n)$ be a deduction from $\Si$. Then  there exists an extension $r'$ of $r$ to $\at(\Si\cup\{\si_1, \ldots ,\si_n\})$ such that $r'\models \Si\cup\{\si_1, \ldots ,\si_n\}$.
\end{lemma}
\begin{proof}
We prove the claim by induction on $n$. We denote by $R_n$ the set $\at(\Si\cup\{\si_1, \ldots ,\si_n\})$. Assuming the claim for $n-1$, we first find an extension $r_{n-1}$ of $r$ to $R_{n-1}$ such that $r_{n-1}\models \Si\cup\{\si_1, \ldots ,\si_{n-1}\}$. If $\si_n$ is obtained by an application of a conjunction or some ind rule, then it is easy to see that we may choose $r_n:=r_{n-1}$. Hence, it suffices to consider the cases where $\si_n$ is obtained by using one of the chase rules. Due to Lemma \ref{clear}, it suffices to find an extension $r_n$ of $r_{n-1}$ to $R_n$ such that $r_n\models \si_n$. In the following cases, $\tuple A$ denotes a sequence listing the attributes of $R\sub R_{n-1}$.

\paragraph{\textbf{Case [CS].}} Assume that $\si_n$ is obtained by [CS] and is of the form $$(T^*,\id)[RS]\ja \bigwedge_{t\in T}  t(\tuple A) \sub  \tuple A$$
where $T=\re{T^*}{R}$, $S=\Val(T)$ consists of \emph{new attributes} and $R$ of \emph{distinct values}. Let $r_n:= r_{n-1}\bo r$ be an extension of $r_{n-1} $ to $R_n=R_{n-1} S$, where 
$$r:= \{ h: h\textrm{ is a valuation on }T \textrm{ such that } h(T) \sub \re{r_{n-1}}{R}\}.$$ We claim that $r_n\models \si_{n}$. Consider the first conjunct of $\si_n$, and let $h$ be a valuation on $T^*$ such that $h(T^*) \sub \re{r_n}{RS}$. Then $\re{h}{S}$ is is a valuation on $T$ such that $h(T) \sub \re{r_{n}}{R}=\re{r_{n-1}}{R}$, i.e., $\re{h}{S}=\re{t_0}{S}$ for some $t_0\in r_n$. Since $R$ consists of \emph{distinct values} and thus $R\cap \Dom(h) =\emptyset$, we may define $h'$ as an extension of $h$ with $A\mapsto t_0(A)$, for $A\in R$. Then $\re{h'}{RS}=\re{t_0}{RS}\in \re{r_n}{RS}$, and therefore $r_n \models (T^*,\id)[RS]$.

Consider then $ t(\tuple A) \sub  \tuple A$, for $t\in T$, and let $t_0\in r_n$.
By the definition, $\re{t_0}{S}=  h$ for some valuation $h$ on $T$ such that   $h(T) \sub \re{r_{n}}{R}$, and hence we obtain that $t_0\circ t(\tuple A)=h\circ t(\tuple A)= t_1(\tuple A)$ for some $t_1\in r_n$. Therefore, $r_n\models  t(\tuple A) \sub  \tuple A$.

\paragraph{\textbf{Case [CR].}} Assume that $\si_n$ is of the form (i) $\bigwedge_{t'\in T'}f\circ t'(\tuple A) \sub \tuple A$ or (ii) $f(x)=f(y)$, and is obtained by [CR] from 

\begin{enumerate}[(i)]
\item $( T, T')[R] \ja \bigwedge_{t\in T}f\circ  t(\tuple A) \sub \tuple  A, $
\item $(T,x=y)[R] \ja \bigwedge_{t\in T} f\circ  t(\tuple A) \sub \tuple A, $
\end{enumerate}
where in case (ii) $f$ is a valuation on $T\cup T'$ such that it is 1-1 on  $S:=\Val(T')\setminus\Val(T)$ and $f(x)$ is a \emph{new attribute} for $x\in S$.  %Let $r_n:= r_{n-1}\bo r$ where $r:= \{h\circ \re{f^{-1}}{S}: h(T') \sub \re{r_{n-1}}{R}\}$,  and consider an ind $f\circ t'(\tuple A) \sub \tuple A$ for 	$t'\in T'$.  
 Let $s\in r_{n-1}$. Since $r_{n-1}\models \bigwedge_{t\in T}f\circ  t(\tuple A) \sub \tuple  A $,   we first obtain that $s\circ f (T)\sub \re{r_{n-1}}{R}$. % for all $s\in r_{n-1}$. 
\begin{enumerate}[(i)]
\item  Since $r_{n-1} \models ( T, T')[R]$ we find a mapping $g:S\to \Val$  such that $h(T')\sub \re{r_{n-1}}{R}$, for $h= g\cup (s\circ f)$. Since $f$ is 1-1 on $S$, we  can now define $r_n$ as the relation obtained from $r_{n-1}$ by extending each $s\in r_{n-1}$ with $f(x) \mapsto g(x)$ for $x \in S$. Then for each $s\in r_n$,  $s\circ f(T')\sub \re{r_n}{R}$, and hence  we obtain that %$s\circ t'(\tuple A) =s'(\tuple A)$ for some $s'\in r_{n-1}$. Therefore 
$r_n \models \bigwedge_{t'\in T'} f\circ t'(\tuple A) \sub \tuple A.$
\item  It suffices to show that $r_{n-1}\models f(x)=f(y)$. Since $s\circ f (x)=s\circ f(y)$ by $r_{n-1} \models (T,x=y)[R]$, this follows immediately.
\end{enumerate}

\paragraph{\textbf{Case [CT].}} Assume that $\si_n$ is of the form (i) $(T ,T')[ R]$ or (ii) $(T ,x=y)[ R]$ and  is obtained by [CT] from 
\begin{enumerate}[(i)]
\item $(T^*,\id)[R S]\ja\bigwedge_{t'\in T'} u\circ t'(\tuple A)\sub \tuple A,$ where  $u$ is a mapping  $\Val(T')\to \at$ that is the identity on $\Val(T)\cap\Val(T')$,
\item $(T^*,\id)[ R  S]\ja x=y,$ where $x,y\in \Val(T)$.
\end{enumerate}
Moreover, in both cases $T=\re{T^*}{R}$, $S= \Val(T)$,  and $\Val(\re{T^*}{S})$ consists of \emph{distinct values}. It suffices to show that $r_{n-1}\models \si_n$, so let $h$ be a valuation on $T$ such that $h(T)\sub \re{r_{n-1}}{R}$. Since $\Val(\re{T^*}{S})$ consists of disctinct values, $h$ can be extended to a valuation $h'$ on $T^*$ such that $h'(T^*)\sub \re{r_{n-1}}{RS}$. Since $r_{n-1}\models (T^*,\id)[R S]$, there is an extension $h''$ of $h'$ to attributes in $R$ such that $\re{h''}{RS} \in \re{r_{n-1}}{RS}$. % and $h''$ assigns new values only to elements of $R$. 
Hence, we obtain that $\re{h}{S} \in \re{r_{n-1}}{S}$. Let then $s\in r_{n-1}$ be such that it agrees with $h$ on $S$.
\begin{enumerate}[(i)]

\item  Since $ r_{n-1}\models \bigwedge_{t'\in T'} u\circ t'(\tuple A)\sub \tuple A$,  we obtain that $s\circ u(T') \sub \re{r_{n-1}}{R}$. Moreover, we notice that $s\circ u = h$ on $\Val(T) \cap \Val(T')$. 

\item Since  $ r_{n-1}\models x=y$, we obtain that $s(x)=s(y)$. Then $h(x)=h(y)$ since  $x,y\in S$. 

\end{enumerate}
Hence, in both cases we obtain that $r_{n-1}\models \si_n$. This concludes the [CT] case and the proof.\qed

\end{proof}

Using the previous lemma,  soundness of the rules follows.

\begin{theorem}
Let $\Si\cup\{\si\}$ be a finite set of egd's and tgd's over $R$. Then $\Si \models \si$ if $\Si \vdash \si$.
\end{theorem}
\begin{proof}
Let $r$ be a relation such that $r\models \Si$, and assume that $(\si_1, \ldots ,\si_n)$ is a deduction from $\Si$ where $\si=\si_n$ contains no attributes that appear as new in $\si_1, \ldots ,\si_n$.  If $R':=\at(\Si\cup\{\si_1, \ldots ,\si_n\})$, then by Lemma   \ref{soundlem}  we find an extension $r'$ of $\re{r}{R}$ to $R'$ such that $r'\models \si$. Then using Lemma \ref{clear} we obtain that $r\models \si$.\qed
\end{proof}

\section{Chase Revisited}\label{chasing}
In this section we define the chase for the class of egd's and tgd's. The chase algorithm was generalized to typed egd's and tgd's in \cite{beeri84}, and here we present the chase using notation similar to that in \cite{abiteboul95}. First let us assume, for notational convenience,  that there is a total, well-founded order $<$ on the set $\Val$, e.g., $x_1<x_2<x_3<\ldots $ for $\Val=\{x_1, x_2, x_3,\ldots  \}$. Let $\Si\cup\{\si\}$ be a set of egd's and tgd's over $R$. % where $\si$ is either $(T,T')$ or $(T,x=y)$.  
 A \emph{chasing sequence} of $\si$ over $\Si$ is a (possibly infinite) sequence $\si_0,\si_1, \ldots ,\si_n,\ldots $ where 
$\si_0=\si$, and
$\si_{n+1}$ is obtained from $\si_n$, with $T:=\pr_1(\si_n)$, according to either of the following rules.

Let $\tau\in \Si$ be of the form $(S,x=y)$, and  suppose that there is a valuation $f$ on $S$ such that $f(S)\sub T$ but $f(x) \neq f(y)$. Then $\tau$ (and $f$) can be applied to $\si_n$ as follows:
\begin{itemize}
\item \textbf{egd rule:}  Let $\si_{n+1}:= g(\si_n)$ where $g: \Val \to \Val$ is the identity everywhere except that it maps $f(y)$ to $f(x)$ if $f(x) < f(y)$, and $f(x)$ to $f(y)$ if $f(y) < f(x)$.
\end{itemize}
Let $\tau\in \Si$ be of the form $(S, S')$, and suppose that there is a valuation $f$ on $S$ such that $f(S) \sub T$, but there exists no extension $f'$ of $f$ to $S'$ such that $f(S')\sub T$. Then $\tau$  can be applied to $\si_n$ as follows:
\begin{itemize}
\item \textbf{tgd rule:}  List all $f_1, \ldots , f_n$ that have the above property, and for each $f_i$ choose a \emph{distinct extension} to $S'$, i.e., an extension $f'_i$ to $S'$ such that each variable in $\Val(S') \setminus \Val(S)$ is assigned a distinct new value greater than any value in $\Val(\si_0)\cup \ldots\cup\Val(\si_n)$. Moreover, no new value is assigned by two $f'_i,f'_j$ where $i \neq j$. Then we let $\si_{n+1}:( T \cup f'_1(S')\cup  \ldots \cup f'_m(S'),\pr_2(\si_{n}))$.
\end{itemize}

Construction of a chasing sequence is restricted with the following two conditions:
\begin{enumerate}[(i)]
\item Whenever an egd is applied, it is applied repeatedly until it is no longer applicable.
\item No dependency is starved, i.e., each dependency that is applicable infinitely many times is applied infinitely many times.
\end{enumerate}
Let $\chase{\Si,\si}=\si_0, \si_1,\ldots $ be a chasing sequence of $\si$ over $\Si$. Due to the possibility of applying egd's, a chasing sequence may not be monotone with respect to $\sub$. Hence, depending on whether $\si$ is a tgd or an egd, we define
\begin{itemize}
\item egd: $\ochase{\Si,\si}:=(T^1,x=y)$,
\item tgd: $\ochase{\Si,\si}:=(T^1,T^2)$,
\end{itemize}
where $T^i:= \{u:\exists m \forall n\geq m (u\in \pr_i(\si_n))\}$ and $x=y$ is $\pr_2(\si_n)$ for $n\in \N $ such that $\pr_2(\si_n)=\pr_2(\si_m)$ for all $m\geq n$. Note that ``newer" values introduced by the tgd rule are always greater than the ``older" ones, and values may only be replaced  with smaller ones. Hence, no value can change infinitely often, and therefore $\ochase{\Si,\si}$ is always well defined and non-empty.

We also associate each chasing sequence with the following descending valuations $\rho_n$, for $n \geq 0$. We let  $\rho_0=\id$, $\rho_{n+1}= g\circ \rho_n$ if $\si_{n+1}$ is obtained by an application of the egd rule where $\si_{n+1}=g(\si_{n})$, and $\rho_{n+1}= \id\circ \rho_n$ otherwise. We then define $\rho(x) = \lim_{n\to \infty} \rho_n(x)$, i.e., $\rho(x) = \rho_n(x)$ if $n\in\N$ such that $\rho_{m}(x) =\rho_n(x)$ for all $m\geq n$. Then we obtain that 
$$\ochase{\Si,\si} =\bigcup_{n=0}^{\infty}\rho(\si_n).$$

A dependency $\tau$ is \emph{trivial} if
\begin{itemize}
\item $\tau$ is of the form $(T,x=x)$, or
\item $\tau$ is of the form $(T,T')$ and there is a valuation $f$ on $T'$ such that $f$ is the identity on $\Val(T)\cap\Val(T')$ and $f(T') \sub T$.
\end{itemize}
%Note that if $\tau$ is full, then the latter condition indicates that $T' \sub T$.

It is well-known that the chase algorithm captures unrestricted implication of dependencies. The proof of the following proposition is hence located in Appendix.
\begin{proposition}\label{chase}
Let $\Si\cup\{\si\}$ be a set of egd's and tgd's over  $R$. Then the following are equivalent:
\begin{enumerate}[(i)]
\item $\Si \models \si$,
\item there is a chasing sequence $\chase{\Si,\si}=\si_0, \si_1,\ldots $ of $\si$ over $\Si$ such that  $\ochase{\Si,\si}$ is trivial,
\item there is a chasing sequence $\chase{\Si,\si}=\si_0, \si_1,\ldots $ of $\si$ over $\Si$ such that  $\si_n$ is trivial, for some $n$.
\end{enumerate}
\end{proposition}
%\begin{proof}
%The if direction follows from an easy  induction  on $i$ showing that, given a relation $r$ with $r\models \Si$, a chasing sequence $\mathcal{T}=T_0, T_1,\ldots $, and a valuation $g$ that embeds $T$ to $r$, then for all $i$ there is an extension of $g$ in $T_i$ to $r$.

%For the only-if direction assume that $\mathcal{T}=T_0, T_1,\ldots $ is an infinite chasing sequence that does not satisfy the condition. Then clearly $\chase(\mathcal{T})\not\models \si$. Also one can show that $\chase(\mathcal{T})\models \Si$ (see the proof of Lemma 12 in [Beeri and Vardi] where this is showed for typed dependencies, and note that it applies to the class of untyped dependencies also).
%\end{proof}

\section{Completeness Theorem}
In this section we show that the rules presented in Definition \ref{axioms} are complete for the implication problem of embedded dependencies. Let us first illustrate the use of the axioms in the following simple example.
\begin{ex}
Consider the implication problem $\{\si,\si'\}\models \tau$ where $\si,\si',\tau $ are illustrated in Fig. \ref{A}, e.g., $\si =(T,t)$ where $T$ consists of the top two rows of $\si$ and $t$ is the bottom row. Note that $\si$ and $\tau$ are embedded multivalued dependencies of the form $A\twoheadrightarrow B|C$ and $A \twoheadrightarrow B|CD$, respectively, and $\si'$ is a functional dependency of the form $C\rightarrow D$.
\begin{figure}[h]
\begin{center}
$ \scalebox{1.2}[1.1]{$\si=$ \begin{tabular}{  |  C{4mm} C{4mm} C{4mm} C{4mm} | }
$A$& $B$ & $C$ & $D$ \\\hline
$a_0$ & $b_0 $ & $c_0$ & $d_0$  \\
$a_0$ &$b_1 $ & $c_1$ & $d_1$\\\hline
$a_0$ &$ b_0$ & $c_1$ & $d_2$ \\

\end{tabular}

\qquad

$\si'= $ \begin{tabular}{  | C{4mm} C{4mm} C{4mm} C{4mm} | }
$A$& $B$ & $C$ & $D$ \\\hline
$a_0$ & $b_0 $ & $c_0$ & $d_0$  \\
$a_1$ &$b_1 $ & $c_0$ & $d_1$\\\hline
   & $d_0$ &$=$ &$d_1$ \\

\end{tabular}
\qquad

$\tau= $ \begin{tabular}{  |  C{4mm} C{4mm} C{4mm} C{4mm} | }
$A$& $B$ & $C$ & $D$ \\\hline
$a_0$ & $b_0 $ & $c_0$ & $d_0$  \\
$a_0$ &$b_1$ & $c_1$ & $d_1$\\\hline
$a_0$ &$ b_0$ & $c_1$ & $d_1$ \\

\end{tabular}

}$\caption{\label{A}}

\end{center}
\end{figure}
It is easy to see that the implication holds, and this can be also verified by a chasing sequence $\tau_0,\tau_1,\tau_2$ of $\tau$ over $\{\si,\si'\}$ where $\tau_2$ is trivial. In the chasing sequence, $\tau_0=\tau$ and $\tau_1$ is the result of applying $\si$ to $\tau_0$. For this, note that there exists two valuations on $T$  that embed $T$ to $\pr_1(\tau_0)$ but has no extension that embeds $t$ into $\pr_1(\tau_0)$. These valuations are the identity and the function $f$ that swaps the values of the top and bottom row of $T$. Then $\tau_1$ is obtained by adding to $\pr_1(\tau_0)$ $\id^*(t)$ and $f^*(t)$ where $\id^*$ and $f^*$
are distinct extensions of $\id$ and $f$ to $t$, e.g., $\id^*=\id$ also on $d_2$ and $f^*$ maps $d_2$ to $d_3$. Also, $\tau_2$ is the result of applying $\si'$ to $\tau_1$ two times, i.e., $\tau_2$ is obtained from $\tau_1$ by replacing $d_3$ with $d_0$ and $d_2$ with $d_1$. Clearly $\tau_2$ is trivial, and hence we obtain the claim by Proposition \ref{chase}.

\begin{figure}[h]
\begin{center}
$ \scalebox{1.2}[1.2]{
$\tau_0= $ \begin{tabular}{  |  C{4mm} C{4mm} C{4mm} C{4mm} | }
$A$& $B$ & $C$ & $D$ \\\hline
$a_0$ & $b_0 $ & $c_0$ & $d_0$  \\
$a_0$ &$b_1$ & $c_1$ & $d_1$\\\hline
$a_0$ &$ b_0$ & $c_1$ & $d_1$ \\

\end{tabular}
\qquad
$\tau_1=$ \begin{tabular}{  |  C{4mm} C{4mm} C{4mm} C{4mm} | }
$A$& $B$ & $C$ & $D$ \\\hline
$a_0$ & $b_0 $ & $c_0$ & $d_0$  \\
$a_0$ &$b_1 $ & $c_1$ & $d_1$\\\hdashline
$a_0$ & $b_0 $ & $c_1$ & $d_2$  \\
$a_0$ &$b_1 $ & $c_0$ & $d_3$\\\hline
$a_0$ &$ b_0$ & $c_1$ & $d_1$ \\

\end{tabular}

\qquad

$\tau_2=$ \begin{tabular}{  |  C{4mm} C{4mm} C{4mm} C{4mm} | }
$A$& $B$ & $C$ & $D$ \\\hline
$a_0$ & $b_0 $ & $c_0$ & $d_0$  \\
$a_0$ &$b_1 $ & $c_1$ & $d_1$\\
$a_0$ & $b_0 $ & $c_1$ & $d_1$  \\
$a_0$ &$b_1 $ & $c_0$ & $d_0$\\\hline
$a_0$ &$ b_0$ & $c_1$ & $d_1$ \\

\end{tabular}

}$\caption{\label{B}}

\end{center}
\end{figure}
This procedure can now be simulated with our axioms as follows. First, with one application of [CS] we derive 
$$(T,\id)[RS] \ja a_0b_0c_0d_0\sub ABCD \ja a_0b_1c_1d_1\sub ABCD$$ 
where $T=\{t,t'\}$, $R=\{A,B,C,D\}$, and $S=\{a_0,b_0,b_1,c_0,c_1,d_0,d_1\}$ is a set of values that are interpreted as new attributes. Here $t(x)$ and $t'(x)$, for $x\in S$, and $A,B,C,D$ are interpreted as  distinct values. $(T,t)[RS]$ is  illustrated in Fig. \ref{C} where all the distinct values are hidden.
\begin{figure}[h]
\center\scalebox{1.2}[1.1]{
$\begin{tabular}{c |  C{4mm} C{4mm} C{4mm} C{4mm} :  C{4mm} C{4mm} C{4mm} C{4mm}C{4mm} C{4mm} C{4mm}  |}
&$A$& $B$ & $C$ &$D$& $a_0$&$b_0$&$b_1$&$c_0$&$c_1$&$d_0$&$d_1$\\\cline{2-12}
$t$&$a_0$ & $b_0 $ & $c_0$ & $d_0$&&&&&&&\\
$t'$&$a_0$ &$b_1 $ & $c_1$ & $d_1$    &&&&&&&\\\cline{2-12}
$\id$&&&&&$a_0$&$b_0$&$b_1$&$c_0$&$c_1$&$d_0$&$d_1$\\\cline{2-12}
\end{tabular}$}\caption{$(T,\id)[RS]$\label{C}}
\end{figure}
Now with one application of [CR], letting $f= \id$, we derive $a_0b_0c_1d_2\sub ABCD$ from 
\begin{equation}\label{this}
\si\ja a_0b_0c_0d_0\sub ABCD \ja a_0b_1c_1d_1\sub ABCD
\end{equation}
Note that in this step, $d_2$ is interpreted as a new attribute. Let then $f$ be the valuation that is the identity on $a_0,b_0,b_1,d_1$, and otherwise maps $a_1\mapsto a_0$, $c_0\mapsto c_1$, and $d_0\mapsto d_2$. We notice that 
$f(a_0b_0c_0d_0)=a_0b_0c_1d_2$ and $f(a_1b_1c_0d_1)=a_0b_1c_1d_1$. Hence, we may derive with one application  of [CR] $f(d_0)=f(d_1)$, i.e., $d_2=d_1$ from 
\begin{equation*}\label{that}
\si' \ja f(a_0b_0c_0d_0)\sub ABCD \ja f(a_1b_1c_0d_1)\sub ABCD.
\end{equation*}
Then we apply [EE] and derive $a_0b_0c_1d_1\sub ABCD$ from
$$d_2=d_1 \ja a_0b_0c_1d_2\sub ABCD$$
Finally, we may apply [CT] and derive $\tau$ from
$(T,\id)[RS] \ja a_0b_0c_1d_1\sub ABCD.$

\end{ex}

The following lemma shows that the above technique extends to all chasing sequences. The proof is straightforward and located in Appendix. %Let $\Si \cup \{\si\}$ be a set of egd's and tgd's over $R$, and assume that $\Si \models \si$. For a deduction of $\si$ from $\Si$, the idea is simulate a chasing sequence  $\chase{\Si,\si}=\si_0,\si_1,\ldots$ so that for any rows $t\in \si_i$, we can derive the inclusion atom $t(\tuple A) \sub \tuple A$ where $\tuple A$ lists $R$ and $t(\tuple A)$ lists values that are here interpreted  as attributes. Once a trivial $\si_n$ is found, then we can derive $\si$ with one application of [CT]. 
\begin{lemma}\label{complem}
Let $\chase{\Si,\si}=\si_0,\si_1,\ldots$ be a chasing sequence of $\si$ over $\Si$, where   $\Si \cup \{\si\}$ is a finite set of egd's and tgd's over $R$, let $\tuple A$ be a sequence listing the attributes of $R$, let $T:=\pr_1(\si)$ and $T_i := \pr_1(\si_i)$, and let $n\in \mathbb{N}$. Then there exists a deduction from $\Si$, with  attributes from $R\cup\bigcup_{i\in \N} \Val(T_i)$, listing the following dependencies:
\begin{enumerate}[(i)]
\item $(T^*,\id)[RS]$
where $\re{T^*}{R}=T$, $S=\Val(T)$, and $ \re{T^*}{S}$ consists of distinct values,
%\item $ t^n (A)= t^{n+1} (A)$, for $n < m$, $t^n\in T_n$, and $A \in R$,
\item $f(x)=f(y)$, for each application of  $(S,x=y)$ and  $f$  to $\si_m$, for $m < n$,

\item $ t(\tuple A) \sub \tuple A$, for $t \in T_m$ where  $m\leq n$.
\end{enumerate}
\end{lemma}

With the lemma, we can now show completeness.

\begin{theorem}
Let $\Si\cup\{\si\}$ be a finite set of egd's and tgd's over $R$. Then  $\Si \models \si \Leftrightarrow \Si  \vdash \si$.
\end{theorem}
\begin{proof}
Assume that $\Si \models \si$, and let $\tuple A$ be a sequence listing $R$. Then by Proposition \ref{chase} there is a chasing sequence $\chase{\Si,\si}=\si_0, \si_1, \ldots $ of $\si$ over $\Si$ such that $\si_n$ is trivial for some $n$. Let $D=(\tau_1, \ldots ,\tau_l)$ be a deduction from $\Si$ obtained by Lemma \ref{complem}, and let $T:=\pr_1(\si)$ and $T_i:=\pr_1(\si_i)$.

Assume first that $\si$ is an egd of the form $(T,x=y)$. Then $\si_n$ is $(T_n,z=z)$ where $z=\rho_n(x)=\rho_n(y)$. % and $z=g_n\circ \ldots \circ g_1 (y)$ for $g_i$ such that $g_i=\id$ if $\si_i$ is obtained by applying the tgd rule, and otherwise $g_i=g$ for the valuation $g$ such that $T_i=g(T_{i-1})$.  
Now, either $\rho_{i+1}(x)$ is $\rho_i(x)$, or the equality $\rho_{i+1}(x)=\rho_i(x)$ (or its reverse) is listed in $D$ by item (ii). Hence, using repeatedly [ES,ET] we may further on derive $z=x$. Since $z=y$ is derivable analogously, we therefore obtain $x=y$ by [ES,ET]. Then with one application of [CT], we  derive $(T,x=y)$ from $(T^*,\id)[RS] \wedge x=y$ where $\re{T^*}{R}=T$. Note that the $(T^*,\id)[RS]$ of the correct form is listed in $D$ by item (i) of Lemma \ref{complem}.

Assume then that $\si$ is a tgd of the form $(T,T')$, and let $ T'_i:= \pr_2(\si_i)$. Then $\si_n$ is $(T_n,T'_n)$, and there is a valuation $f$ on $T_n'$ such that $f$ is the identity on $\Val(T_n)\cap\Val(T'_n)$ and $f(T'_n)\sub T_n$.  Let  $t'\in T'$. Then $\rho_n\circ t'\in T'_n$ and by item (iii) of Lemma \ref{complem} we obtain that $f\circ \rho_n\circ t'(\tuple A )\sub \tuple A$ is listed in $D$.
%Note that $T'_n= h (T')$ where
 %Let
 %$h=g_n\circ \ldots \circ g_1$ be  the sequence of valuations $g_1, \ldots ,g_n$ defined in the previous paragraph, %Hence $f\circ h\circ t' \in S$.
For $A\in R$, we  have then two cases :
\begin{itemize}
\item If $t'(A) \in \Val(T')\cap\Val(T)$, then we first notice that $f\circ \rho_n\circ t'(A)$ is $\rho_n\circ t'(A)$ since  $\rho_n\circ t'(A) \in  \Val(T'_n)\cap\Val(T_n)$. Also we notice that the equality $\rho_n\circ t'(A)=t'(A)$ can be derived analogously to the egd case. 
\item If $t'(A) \in \Val(T')\setminus \Val(T)$, then $f\circ \rho_n \circ t' (A)= f\circ t'(A)$ since by the definition of the chase $\rho_n$ is the identity on $\Val(T')\setminus\Val(T)$. 

\end{itemize}
Now, letting $f^*$ be the mapping $\Val(T')\to \at$ which is the identity on $\Val(T')\cap \Val(T)$ and agrees with $f$ on $\Val(T')\setminus\Val(T)$, we can by the previous reasoning and using repeatedly [EE]  derive $f^*\circ t'(\tuple A) \sub \tuple A$ from $f\circ \rho_n \circ t' (\tuple A) \sub \tuple A$. Finally, we can then with one application of [CT] derive $(T,T')$ from 
$$(T^*,\id)[RS] \ja \bigwedge_{t'\in T'} f^*\circ t'(\tuple A) \sub \tuple A.$$ \qed

\end{proof}

\section{Typed dependencies}
Consider then the class of typed embedded dependencies. In this setting [CS] and [CT] can be replaced with rules that involve only embedded join dependencies (ejd's) and inclusion dependencies. %A team $X$ satisfies a join atom if the join of some of its restrictions can be extended to it.
We define ejd's over tuples of attributes as follows.
\begin{definition}\label{joinatom}
Let $\tuple A_1, \ldots ,\tuple A_n$ be tuples of attributes listing $R_1, \ldots ,R_n$, respectively, and let $R:= \bigcup_{i=1}^n R_i$. Then $\bow{\tuple A_i}_{i=1}^n$ is an \emph{embedded join dependency} with the semantic rule
\begin{itemize}
\item $r \models\hspace{1mm} \bow{\tuple A_i}_{i=1}^n$ if and only if $\re{r}{ \tuple R} = \re{r}{  \tuple R_1}\bo \ldots \bo\re{r}{\tuple R_n}$.
\end{itemize}

\end{definition}
The two alternative rules for the chase are now the following. We call a relation  typed if none of its values appears in two distinct columns. %In the rules, we denote by $\tuple A \approx \tuple B$ the conjunction $\tuple A \sub \tuple B \ja \tuple B \sub \tuple A$.

%It is straightforward to see that for typed egd's and tgd's we obtain a similar complete axiomatization by replacing [CT] with [CT*] and [CR] with [CR*]. In the rules $\tuple A$ is sequence listing the attributes of $R$.

%\begin{itemize}
%\item[CT*] Chase Termination: 
%$$\textrm{if } \bigwedge_{i=1}^m  \big (\tuple A\sub \tuple B_i  \ja \tuple B_i = s_i(\tuple A)\big ) \ja \bo(B_i)_{i=1}^m \ja \bigwedge_{i=1}^{m'}t_i(\tuple A) \sub  \tuple A, \textrm{ then }\sigma
%$$
%where $\sigma[\tuple A] = (s_1, \ldots ,s_m, t_1, \ldots ,t_{m'})\in \mathcal{C}^*$, and  $\tuple B_i$ are pairwise disjoint sequences.
%\end{itemize}

%\textbf{Rules for EJDs}:
\begin{itemize}
\item[CS*] Chase Start$^*$: $$\bigwedge_{t\in T }\tuple A \sub t(\tuple A) \wedge  \bow{t(\tuple A)}_{t\in T} \wedge \bigwedge_{t\in T } t(\tuple A) \sub \tuple A  %\ja \bo (\tuple A,\tuple B)
$$
 where $T$ is a typed relation and $\Val(T)$ is a set of \emph{new attributes}.

%$\tuple B= \tuple B_1 \ldots \tuple B_n =B_{i_1} \ldots B_{i_m} $ renames $\tuple A = \tuple A_1 \ldots \tuple A_n =A_{j_1} \ldots A_{j_m}$ (i.e. $j_k=j_l \Leftrightarrow i_k = i_l$, for $1 \leq k < l \leq m$) with \emph{new} attributes.

\item[CT*] Chase Termination$^*$: 
$$tgd: \textrm{if } \bigwedge_{t\in T}  \tuple A\sub t(\tuple A)  \ja \bow{t(\tuple A)}_{t\in T} \ja \bigwedge_{t'\in T'}u\circ t'(\tuple A) \sub  \tuple A, \textrm{ then }(T,T')[R],
$$
$$egd: \textrm{if } \bigwedge_{t\in T}  \tuple A\sub t(\tuple A)  \ja \bow{t(\tuple A)}_{t\in T} \ja x=y, \textrm{ then }(T,x=y)[R],
$$
where tgd: $u$ is a mapping  $\Val(T')\to \at$ that is the identity on $\Val(T')\cap\Val(T')$, and egd: $x,y\in \Val(T)$.
\end{itemize}

The first rule is sound for typed dependencies since, for arbitrary $r$ with $\Dom(r)\cap \Val(T) =\emptyset$,  an instance of [CS*] is satisfied by $r\bo q(r)$ where $q$ is the SPJR query
$$ \rho_{t_1(\tuple A)/\tuple A} \tuple A \bo \ldots \bo  \rho_{t_n(\tuple A)/\tuple A} \tuple A,$$
where $\rho$ is the rename operator and $T=\{t_1, \ldots ,t_n\}$. However, a counter example  for soundness can be easily constructed for untyped dependencies. If $T$ and $r$ are the relations illustrated in Fig. \ref{loppu}, then  no extension $r'$ of $r$ to  $\Val(T)$ satisfies $\bigwedge_{t\in T} t(AB)\sub AB$.
\begin{figure}[h]
\begin{center}
$ \scalebox{1.2}[1.1]{

$T= $ \begin{tabular}{ c | C{3.5mm} C{3.5mm} C{3.5mm}  | }
&$A$& $B$ \\\hline
$t$&$x$ & $y $  \\
$t'$&$y$ &$x $ \\\hline

\end{tabular}

\qquad

$r=$ \begin{tabular}{ c | C{3.5mm} C{3.5mm} | }
&$A$& $B$ \\\hline
$s$&$0$ & $1 $  \\\hline
\end{tabular}

%\qquad

%$r'= $ \begin{tabular}{ c | C{3.5mm} C{3.5mm} C{3.5mm} C{3.5mm}  | }
%&$A$& $B$ & $x$ &$y$\\\hline
%$s$&$0$ & $1 $ & $0$&$1$ \\\hline
%$s'$&$0$ &$1 $ & $1$&$0$\\\hline

%\end{tabular}
}$\caption{\label{loppu}}

\end{center}
\end{figure}

 Soundness of [CT*] is obtained analogously to that of [CT]. Also, completeness is obtained by deriving exactly in the same way as in the general case,   $\bigwedge_{t'\in T'}u\circ t'(\tuple A) \sub  \tuple A$ (in the tgd case) or  $x=y$ (in the egd case)  from $\bigwedge_{t\in T} t(\tuple A) \sub \tuple A$.  Let us then write $\Si \vdash^*\si $ if $\si$ is deduced from $\Si$ in the sense of Definition \ref{dedu} and using  rules [EE,CS*,CR,CT*] together with elimination and introduction of conjunction. Then we obtain the following theorem.
\begin{theorem}
Let $\Si\cup\{\si\}$ be a finite set of typed egd's and tgd's over $R$. Then $\Si \models \si \Leftrightarrow \Si \vdash^*\si$.
\end{theorem}

\section*{Acknowledgement}
The author was supported by grant 264917 of the Academy of Finland.
\bibliographystyle{splncs}
\bibliography{biblio}
\newpage
%\section{Conclusion}
\section*{Appendix}

\begin{proof}(Proposition \ref{chase})
Let $\Si\cup\{\si\}$ be a set of egd's and tgd's over $R$. The direction 
$(ii)\Rightarrow (iii)$ is clear because it suffices to choose $\si_n$ such that all the relevant tuples and values remain fixed in $\si_{m}$ for $m\geq n$. We show $(i)\Rightarrow (ii)$ and $(iii)\Rightarrow (i)$.

$(i)\Rightarrow (ii)$: Assume that there is a chasing sequence $\chase{\Si,\si}=\si_0, \si_1,\ldots $ of $\si$ over $\Si$ such that  $\ochase{\Si,\si}$ is non-trivial. We claim that $\ochase{\Si,\si}\models \Si$ and $\ochase{\Si,\si} \not\models \si$. Let $T_n$  denote $\pr_1(\si_n)$. Assume first that $(S,x=y)\in \Si$ and assume to the contrary that $f$ is a valuation such that $f(S) \sub \ochase{\Si,\si}$ but $f(x)\neq f(y)$. Then there exists $m \in \N$ such that  $f(S)\in T_n$ for all $n\geq m$, contradicting the assumption that no dependency is starved in the chase. 

Assume that $(S,S')\in \Si$, and assume that $f$ is a valuation such that $f(S) \sub \ochase{\Si,\si}$, and let $m \in \N$ be such that  $f(S)\sub T_n$ for all $n\geq m$. Then there is an extension $f'$ of $f$ to $S'$ such that $f'(S')\sub T_{m'}$ for some $m'\geq m$, where we define $T'_n:=\pr_2(\si_n)$. % Let $g_i$ be the function associated with the construction of $\chase{\Si,\si}$, i.e., $g_i$ is such that $g_i=\id$ if $\si_i$ is obtained by applying the tgd rule, and otherwise $g_i=g$ for the valuation $g$ such that $T_i=g(T_{i-1})$. 
%Then there exists $n'' \geq n'$ such that, for $g=g_{n''}\circ \ldots \circ g_{n'+1}$, we have that $g\circ f(S') \sub T_n$ for all $n \geq n''$, i.e., $g\circ f  \sub \ochase{\Si,\si}$.
Note that  $\rho_n\circ f'(S) \sub T'_n$ for all $n\geq m'$, and  hence there exists $m''\geq m'$ such that $\rho_{m''}\circ f' (S) \sub T'_{n}$ for all $n \geq m''$.
 %Note that $g$ is the identity on $f(S)$, and hence $g\circ f$ is an extension of $f$. 
Since $f(S) \sub \ochase{\Si,\si}$,  $\rho_{m''}$ is the identity on $f(S)$, and hence we obtain that $\ochase{\Si,\si}\models (S,S')$.

Finally, we show that $\ochase{\Si,\si}\not\models \si$. %If $\si$ is of the form $(T,x=y)$, then a
Analogously to the previous case we find a valuation $\rho_n$ such that $\rho_n(T) \sub \pr_1(\ochase{\Si,\si})$. If $\si$ is of the form $(T,x=y)$, then we obtain that $\rho_n(x)=\rho_n(y)$ is $\pr_2(\ochase{\Si,\si})$. Since $\ochase{\Si,\si}$ is non-trivial, $\rho_n(x)$ and $\rho_n(y)$ must  be two distinct values. Hence, $\rho_n$  witnesses $\ochase{\Si,\si}\not\models (T,x=y)$. 

Assume then that $\si$ is of the form $(T,T')$. Then analogously  $\rho_n\circ T \sub \pr_1(\ochase{\Si,\si})$ and $\rho_n\circ T' = \pr_2(\ochase{\Si,\si})$ for some $n\in\N$. Also note that by the construction $\rho_n$ is the identity on $\Val(T')\setminus\Val(T)$. Now, if there is an extension $h$ of $\re{\rho_n}{\Val(T)}$ to $T'$ such that $h(T')\sub \pr_1(\ochase{\Si,\si})$, then $\ochase{\Si,\si}$ is trivial. Hence $\rho_n$ is a witness of $\ochase{\Si,\si}\not\models \si$.

$(iii)\Rightarrow (i)$: Let $\chase{\Si,\si}=\si_0, \si_1,\ldots $ be a chasing sequence of $\si$ over $\Si$, where $\si_n$ is trivial, and let $T_i$ (or $T$) denote $\pr_1(\si_i)$ (or $\pr_1(\si)$). Assume that $r\models \Si$, and let $f$ be a valuation on $T$ to $r$.  Using the chase construction rules and the assumption it is easy to show  inductively that for all $n$ there is an extension $f_n$ of $f$ to $\cup_{i=0}^n T_i$ such that 
\begin{enumerate}[(i)]
\item $f_n(T_n)\sub r$,
%\item $f_n$ agrees with $f$ on $\Val(T)\cap\Val(T_n)$,
\item  $f_n\circ  \rho_n=f_n$.
\end{enumerate} Assume first that $\si$ is of the form $(T,x=y)$, and hence $\rho_n(x)=\rho_n(y)$. Then by the induction claim we obtain that $f(x)=f(y)$. Assume that $\si$ is of the form $(T,T')$, and let $h$ be a valuation such that $h(T'_n) \sub T_n$ and $h$ is the identity on $\Val(T_n)\cap\Val(T'_n)$. Then $f_n \circ h \circ \rho_n (T') \sub r$ where, by the induction claim, $f_n \circ h \circ \rho_n$ is $f$ on $\Val(T)\cap\Val(T')$. Hence, we obtain that $r\models \si$ in both cases. This concludes the proof.\qed

\end{proof}

\begin{proof}[Lemma \ref{complem}]
W.l.o.g. we may assume that no attribute of $R$ appears as a value in the  chasing sequence, i.e., $R\cap \bigcup_{i\in\N} \Val(\si_i)=\emptyset$. We show the claim by induction on $n$. 
\paragraph{\textbf{The base case.}} First it suffices to deduce by one application of [CS] 
$$(T^*,\id)[RS]\ja \bigwedge_{t\in T}  t(\tuple A) \sub \tuple A$$ where $(T^*,\id)[RS]$ is of the form described in (i).

\paragraph{\textbf{The inductive step.}} Assuming the claim for $n$, we next show the claim for $n+1$.  Assume first that $\si_{n+1}$ is obtained from $\si_n$ by using the egd rule for  $(S,x=y)\in \Si$  %, repeatedly until it is no more applicable.
% W.l.o.g. we may assume that the egd rule is applied only once (otherwise it suffices to repeat the following procedure). %We have two cases, either $\si_m=(T_m,T'_m)$ or $\si_m=(T_m,u=v)$. 
over a valuation $f$ on $S$  such that $f(S)\sub T_n$ and $f(x) \neq f(y)$. Then $T_{n+1}= g(T_n)$ %or $\si_{m+1}:= (g(T_m),g(u)=g(v))$ 
where $g$ is the identity everywhere except that it maps, say $f(y)$ to $f(x)$. 
By the induction assumption, it now suffices to consider  (ii) and (iii) only in the cases that associate with the construction of $\si_{n+1}$. 
\begin{itemize}
\item[(ii)]   The  equality $f(x)=f(y)$ can be derived with one application of [CR], since $f\circ s(\tuple A) \sub \tuple A$, for all $s\in S$, have been deduced by the assumption.
\item[(iii)] Let $t\in T_{n+1}$, and let $t'\in T_{n}$ be such that $t=g\circ t'$. If $f(y)\not\in\Val(t')$, then $t(\tuple A) \sub \tuple A$ has been derived by the induction assumption. Otherwise, $t'(A)=f(y)$ for some $A \in R$. Now  using repeatedly [EE] to $ f(y) = f(x)$ and $t'(\tuple A) \sub \tuple A$, we obtain $t(\tuple A) \sub \tuple A$.
\end{itemize}

Assume then that $\si_{n+1}$ is obtained from $\si_n$ by using the tgd rule for  $(S,S')\in \Si$.  W.l.o.g. we may assume that there is only one valuation  $f$ on $S$ with the property that $f$ embeds $S$ to $T_n$, but no extension of $f$ to $S'$  embeds $S'$ to $T'_n$. Let $f'$ be the distinct extension associated with this step, i.e., $f'$ is an extension of $f$ to $S'$ such that each variable in $\Val(S') \setminus \Val(S)$ is assigned a distinct new value greater than any value appearing in $\si_0, \ldots ,\si_n$. By the induction assumption, none of these new values appear in the deduction. Hence, by the assumption we may with one application of [CR] from $(S,S')\wedge \bigwedge_{s\in S} f'\circ s(\tuple A)$ deduce $\bigwedge_{s'\in S'} f'\circ s' (\tuple A) \sub \tuple A$ where all the new values are interpreted as new attributes. Since $T_{n+1} = T_n\cup f'(S')$, this concludes item (iii) and thus the proof.\qed

\end{proof}

\end{document}